\documentclass[aps,prl,twocolumn]{revtex4}%
\usepackage{graphics}
\usepackage{amsmath}
\usepackage{graphicx}
\usepackage{amsfonts}
\usepackage{amssymb}
\usepackage{float}
\usepackage{longtable}
\usepackage{epsfig}
\usepackage{latexsym}
\usepackage{theorem}
\usepackage{bbm}
\usepackage{bm}
\usepackage{psfrag}

\begin{document}
\title{On quantum illumination, quantum reading,\\and the capacity of quantum computation}
\author{Stefano Pirandola}
\affiliation{Department of Computer Science, University of York, York YO10 5GH, UK}
\affiliation{Research Laboratory of Electronics, Massachusetts Institute of Technology,
Cambridge MA 02139, USA}

\begin{abstract}
In this brief note, I clarify the crucial differences between three different
protocols of quantum channel discrimination, after some confusion has appeared
in recent literature.

\end{abstract}
\maketitle

\subsection{Reason for this note}

In some recent literature, there has been confusion between the
protocols of quantum illumination~\cite{QIll1,Qill2}, quantum
reading~\cite{QR}, and a scheme of communication within a
discrete-variable quantum computer~\cite{Bose}. All these
protocols are based on the idea of quantum channel discrimination
(QCD), but they have completely different applications and
features, which is the reason why they have different names and
should not be confused. Let me provide some clarifications below.

\subsection{Quantum Illumination}

Quantum illumination~\cite{QIll1,Qill2} (see
also~\cite{56,57,58,59,60,61,62,63,64,65,66,67,68,69,70,71,72}) is the use of
input quantum resources (such as entanglement) and output quantum measurements
to enhance the detection of a remote low-reflectivity object in a bright
thermal-noise environment. It can be represented as a QCD problem where the
bit of information associated with the presence or absence of the target is
associated with the binary discrimination of two channels, one including a
partial reflection from the target and the other one being a completely
thermalizing channel (replacing the input with the state of the environment).
Here one can show that, despite initial entanglement is lost in the
sender-receiver path, the benefits of quantum illumination still survive in
the forms of output correlations. These allow one to enhance the sensitivity
of detecting the presence of the target-object with respect to the use of
classical sources of light (in particular separable states in the DV version
of the protocol~\cite{QIll1}, and mixtures of coherent states in the CV
version~\cite{Qill2}). It is called \textquotedblleft
quantum\textquotedblright\ illumination because because it proves a quantum
advantage with respect to classical strategies under the same conditions
(e.g., the same mean number of input photons).

\subsection{Quantum Reading}

Quantum reading~\cite{QR} (see
also~\cite{36,37,38,39,40,41,42,43,44,45,46,47,48,49,50,51}) is the use of
input quantum resources (such as entanglement) and output quantum measurements
to enhance the retrieval of classical information stored in the cells of an
optical memory. It can be represented as a QCD problem where the bit of
information is encoded in two different reflectivities of the memory cell.
These are two different bosonic Gaussian channels that are generally
characterized by different losses and thermal noises. Contrary to quantum
illumination, the scheme is in the very near range, works with high
reflectivities and allows one to use codewords to encode information in blocks
of many cells (so that quantum reading capacities can be defined). Here one
can show that the use of quantum resources (e.g. entanglement) allows one to
enhance the data readout in terms of bits per cell with respect to the use of
classical strategies (in particular the use of coherent states or their
mixtures). It is called \textquotedblleft quantum\textquotedblright\ reading
because it proves a quantum advantage with respect to classical strategies
using the same amount of energy (mean number of photons).

The two schemes of quantum illumination and quantum reading have a specific
peculiarity (quantum enhancement) that gives them the \textquotedblleft
quantum name\textquotedblright. At the same time it is clear that they are
both schemes of QCD, where classical information is retrieved from a box
(target object or memory cell). For instance, see the discussion in
Section~V.H \textquotedblleft Gaussian channel discrimination and
applications\textquotedblright\ of the Gaussian information review~\cite{RMP}.
For more details on these protocols, see also the recent review on photonic
quantum sensing~\cite{REVsensing}.

\subsection{Capacity of quantum computation}

The scheme of Ref.~\cite{Bose} is about the communication capacity of quantum
computation. Clearly, it is not about target detection or optical storage, but
rather communication between registers of a dicrete-variable quantum computer.
In this scheme, there is a \textquotedblleft memory\textquotedblright%
\ register ($M$) where the sender encodes a classical variable $i$ in $N$ pure
quantum states $\left\vert i\right\rangle _{M}\left\langle i\right\vert $ with
some probability $p_{i}$. Then, the receiver has a computation register ($C$)
prepared in some initial state $\rho_{C}^{0}$. The initial state of the two
registers is therefore the tensor-product%
\begin{equation}
\sum_{i}p_{i}\left\vert i\right\rangle _{M}\left\langle i\right\vert
\otimes\rho_{C}^{0}.\label{input}%
\end{equation}
The two registers are then fed into a quantum computer, which applies the
unitary $\hat{U}_{i}$ onto register $C$ conditionally on the value $i$ of
register $M$. Here $\hat{U}_{i}$ represents a series of quantum gates which
describes some quantum algorithm. For instance, $i$ may be an integer, and the
computational output $\rho_{C}^{i}=\hat{U}_{i}\rho_{C}^{0}\hat{U}_{i}^{\dag}$
may be its factorization according to Shor's algorithm.

In general, for the input state as in Eq.~(\ref{input}), the quantum computer
provides the output%
\begin{equation}
\sum_{i}p_{i}\left\vert i\right\rangle _{M}\left\langle i\right\vert
\otimes\rho_{C}^{i}.
\end{equation}
The receiver measures register $C$ so as to discriminate between the possible
output states $\rho_{C}^{i}$, or equivalently between the possible unitary
operations $\hat{U}_{i}$. The optimal information accessible to the receiver
is the Holevo bound%
\begin{equation}
I(C:i)=S(C)-S(C|i),
\end{equation}
where $S(C)$ is the von Neumann entropy of the reduced state of $C$, and
$S(C|i)$ is the corresponding conditional von Neumann entropy. This is clearly
maximized when $p_{i}$ is uniform and the states $\rho_{C}^{i}$ are pure and
orthogonal, so that it takes the maximum value $I(C:i)=\log_{2}N$.

By construction, it is clear that $I(C:i)$ represents the capacity of the
quantum computation $\{\hat{U}_{i}\}$ because it tells you how good the
quantum computer is in providing distinguishable output states (solutions) for
different inputs. When the maximum $\log_{2}N$ is achieved, it means that the
quantum computation is perfect over the entire input alphabet of $N$ letters.

\subsection{Clarifications}

Apart from being interpreted as protocol of QCD, the scheme of
Ref.~\cite{Bose} is clearly different from both quantum illumination and
quantum reading.

\begin{itemize}
\item First of all, Ref.~\cite{Bose} is a communication scheme, where sender's
input alphabet is decoded by a receiver. More specifically, it is
\textit{spatial} communication between two registers which is mediated by a
quantum computation. It is a two-register description where the unitaries are
$\hat{U}_{i}$ are not stored in the computational register $M$ but rather
applied in the dynamical process of the quantum computer (they are in fact
control-unitaries). In this regard it is clearly different from the static
scenario where a classical variable is physically and stably stored into a
black box by an ensemble of channels (to describe presence/absence of a
target, or the different reflectivities of a memory cell). This means that
Ref.~\cite{Bose} is not about readout from \textit{storage}.

\item The input-output process is based on a single signal system processed by
a unitary. Today, we know that unitary discrimination can be perfectly solved
in a finite number of uses~\cite{Acin}. It is therefore different from what
happens in the more general discrimination of quantum channels, where perfect
discrimination is not guaranteed (at finite energies) and the optimal states
may require the use of idler systems, which are not sent through the box but
directly to the output measument in order to assist the entire process.

\item The signal-idler structure, which is missing in Ref.~\cite{Bose}, is one
of the main features for both quantum illumination and quantum reading. The
use of input entanglement and, more generally, quantum correlations, is the
main working mechanism of these two protocols under completely general
conditions of decoherence. As a matter of fact, as already said above, the
\textquotedblleft quantum\textquotedblright\ name in \textquotedblleft
illumination\textquotedblright\ and \textquotedblleft
reading\textquotedblright\ exactly comes from the comparison of a quantum
resource at the input (entanglement) with respect to the use of classical
input states (separable states, mixtures of coherent states).

\item The communication scheme of Ref.~\cite{Bose} is for a discrete-variable
Hilbert space. The main setting for both quantum illumination and quantum
reading is \textit{bosonic}. Quantum illumination provides the possible
working mechanism of a lidar (in the optical case) and a radar (in the
microwave case). Quantum reading is also working at the optical frequencies,
which is the physics in optical storage.
\end{itemize}

In summary, all the protocols of quantum illumination, quantum reading, and
communication of quantum computation can be represented as schemes of QCD.
However, they are protocols with different aims and features, which is the
reason why they should not be confused one with the other. In particular, the
scheme of Ref.~\cite{Bose} is about communication between registers of a
quantum computer, clearly not \textquotedblleft quantum
reading\textquotedblright\ of a classical memory nor \textquotedblleft quantum
illumination\textquotedblright\ of a remote target. With these two protocols,
it only shares the basic structure, that of QCD, which is ultimately about the
retrieval of classical information from some type of black box.

\end{document}